**Formation of Silicon Nanocrystals in Silicon Carbide Using Flash Lamp Annealing**

Charlotte Weiss[1], Manuel Schnabel[1], Slawomir Prucnal[2], Johannes Hofmann[1,] Andreas Reichert[3], Tobias

Fehrenbach[1], Wolfgang Skorupa[2], Stefan Janz[1]

[1]*Fraunhofer Institute for Solar Energy Systems, Heidenhofstraße 2,79110 Freiburg, Germany.*

[2]*Institute of Ion Beam Physics and Materials Research, Helmholtz-Zentrum Dresden-Rossendorf*
*P.O. Box 510119, 01314 Dresden, Germany.*

[3]Now with: *Department of Radiology, Division of Medical Physics, University of Freiburg, Freiburg, Germany*

0)    ABSTRACT

During the formation of Si nanocrystals (Si NC) in $Si_xC_{1-x}$ layers via tube furnace solid-phase crystallization, the unintended formation of nanocrystalline SiC reduces the minority carrier lifetime and therefore the performance of $Si_xC_{1-x}$ as an absorber layer in solar cells. A significant reduction in the annealing time may suppress the crystallization of the SiC matrix while maintaining the formation of Si NC. In this study, we investigated the crystallization of stoichiometric SiC and Si-rich SiC using conventional rapid thermal annealing (RTA) and nonequilibrium millisecond range flash lamp annealing (FLA). The investigated $Si_xC_{1-x}$ films were prepared by plasma-enhanced chemical vapor deposition (PECVD) and annealed at temperatures from 700°C to 1100°C for RTA and at flash energies between 34 J/cm² and 62 J/cm² for FLA. Grazing incidence X-ray diffraction and Fourier transformed infrared spectroscopy were conducted to investigate hydrogen effusion, Si and SiC NC growth and SiC crystallinity. Both the Si content and the choice of the annealing process affect the crystallization behavior. It is shown that under certain conditions, FLA can be successfully utilized for the formation of Si NC in a SiC matrix, which closely resembles Si NC in a SiC matrix achieved by RTA. The samples must have excess Si, and the flash energy should not exceed 40 J/cm² and 47 J/cm² for $Si_{0.63}C_{0.37}$ and $Si_{0.77}C_{0.23}$ samples, respectively. Under these conditions, FLA succeeds in producing Si NC of a given size in less crystalline SiC than RTA does. This result is discussed in terms of nucleation and crystal growth using classical crystallization theory. For FLA and RTA samples, an opposite relationship between NC size and Si content was observed and attributed either to the dependence of H effusion on Si content or to the optical absorption properties of the materials, which also depend on the Si content.

I)      INTRODUCTION

The concept of multijunction solar cells exceeding the Shockley-Queisser limit in efficiency is a success story. III-V Semiconductor cells reach efficiencies of more than 38.8% under one sun [1]. For low-cost Si-based a-Si/μc-Si cells, the highest reported efficiency of 13.6% lies above the corresponding efficiencies of a-Si and μc-Si single junction cells [1]. Between these two record efficiencies, there is still a broad research field of cost reduction in the case of III-V semiconductors and efficiency as well as quality enhancement in the case of Si. One approach is the all-Si tandem solar cell combining a monocrystalline Si bottom cell with a Si nanocrystal (Si NC) top cell [2]. In this structure, the band gap of the top cell has to exceed the band gap of the crystalline Si bottom cell. This condition can be realized with the help of the quantum confinement effect in Si NC, which will occur if the crystals are embedded in a dielectric matrix [3]. 3C-SiC is a suitable matrix material with a band gap of 2.4 eV [4], resulting in relatively small band offsets to Si [5]. Therefore, the material provides the necessary potential well structure for the quantum confinement effect in Si NC on the one hand, while it is sufficiently conductive to act as a solar cell absorber material on the other hand. For this reason, several research groups produce Si NC in a SiC matrix by the deposition of amorphous $Si_xC_{1-x}$ ($x > 1$) layers either by sputtering or by plasma-enhanced chemical vapor deposition (PECVD) [6-11]. During the subsequent annealing step at temperatures up to 1100°C, phase separation between Si and SiC and solid-phase crystallization occur. In the aforementioned studies, two main challenges are reported: the size control of Si NC and the co-crystallization of Si and SiC. The former is crucial for the quantum confinement effect, and the latter causes a high defect density and therefore extremely short lifetimes, making a definitive proof of quantum confinement – to the best of the authors' knowledge – impossible until now. In a prior study, we showed that the Si NC size can be controlled by the Si content of the $Si_xC_{1-x}$ layers [12]. In the present work, we tried to reduce SiC crystallization by reducing the annealing time. Wan *et al.* [13] reported a positive effect on Si crystallization for co-sputtered $Si_xC_{1-x}$ layers by reducing the annealing time from 60 min via furnace annealing to 2 min via rapid thermal annealing (RTA). Flash lamp annealing (FLA), a nonequilibrium thermal process that operates in the millisecond range, has been successfully used to crystallize relatively thick SiC and Si layers. Ohdaira *et al.* [14, 15] used FLA for 10 ms to crystallize 100 µm thick a-Si, whereas Skorupa *et al.* [16] showed that 3C-SiC on Si can be made by FLA for 20 ms. On our PECVD deposited a-$Si_xC_{1-x}$:H layers, we performed an RTA process for 2 min at different temperatures and compared the result to that obtained for samples annealed by FLA for 20 ms at different flash energies. Three different compositions of as-deposited (as-dep) a-$Si_xC_{1-x}$:H layers were investigated: $x = 0.50$, $x = 0.63$ and $x = 0.77$. The crystallization process was studied by grazing incidence X-ray diffraction (GIXRD). We consider the appearance of these patterns to establish a correspondence between FLA energies and RTA temperatures. The diffraction patterns were also used to determine the Si and SiC NC size. With the help of Fourier transformed infrared spectroscopy (FTIR), it was possible to observe the temperature dependence of hydrogen effusion (H effusion) and to compare the SiC crystallinity after RTA and FLA. The effects of layer composition and annealing procedure on the crystallization process are discussed in detail in terms of H effusion, nucleation and crystal growth.

II)     EXPERIMENTAL

## 2.1     *Sample Preparation*

The layers used for RTA and FLA in this study were deposited on 250 µm thick p-type FZ silicon substrates, (100)-oriented with a resistivity of 10 Ω cm. Fused silica (Suprasil®1) substrates measuring 1 mm thick were used for spectrophotometry measurements on as-dep samples. All substrates were cleaned in hot $HNO_3$ and dilute HF before film deposition. The fused silica was subjected to an additional cleaning step in hot $HCl/H_2O_2$ solution followed by a second HF etching step before film deposition.

Film deposition was performed by PECVD in a Roth&Rau AK400 reactor. The substrate temperature was kept at 270°C during deposition and the pressure at 0.3 mbar. The plasma power density was 100 mW/cm², and the plasma frequency was 13.56 MHz. The stoichiometry of the deposited hydrogenated amorphous silicon carbide (a-$Si_xC_{1-x}$:H) was adjusted by varying the gas fluxes of silane ($SiH_4$), methane ($CH_4$) and hydrogen ($H_2$). The three different layer compositions of $x = 0.50$, $x = 0.63$ and $x = 0.77$ used in this work were previously determined by Rutherford backscattering spectrometry with an accuracy of 1% [17]. The associated gas fluxes can be found in our previous work [12].

The intended thickness of the as-dep layers was approximately 200 nm. The deposition rate was confirmed on test samples with the help of spectral ellipsometry. During RTA and FLA, a thickness reduction of nearly 30% occurs because of H effusion and reorganization of the atoms. The resulting thicknesses were measured based on scanning electron microscopy (SEM) images of the layers' cross sections.

## 2.2     *Rapid Thermal Annealing (RTA)*

RTA was conducted in an RTP Sol Invictus 156 PV tool from Centrotherm with a pyrometer for temperature monitoring. The samples were placed on a 4'' FZ Si support wafer, and the front side was heated by the halogen lamp field located above. The annealing temperatures, which varied between 700°C and 1100°C, were reached without a preanneal step at a heating rate of 50°C/s and a plateau time of 2 min for all samples. The entire annealing process was completed under a nitrogen flow of 5 l/min. Only layers on Si substrates were treated by RTA.

### 2.3    *Flash Lamp Annealing (FLA)*

The FLA setup used in this work was described by Prucnal *et al.* [18]. The setup features an RTA system for backside preheating and an FLA system for front-side heating. The RTA system consists of halogen lamps and can reach a maximum temperature of approximately 950°C for silicon wafers. The FLA system is equipped with Xe lamps and can reach temperatures above 1400°C for annealing times between 0.8 and 20 ms. A special reflector is mounted above the Xe lamps to ensure a homogeneous temperature distribution across the sample surface. The spectrum of the Xe lamps used for this experiment has its maximum intensity between 400 and 600 nm. FLA was performed under Ar flow and on films on Si substrates only.

The preanneal step was performed at temperatures between 600 and 800°C for 2, 3 or 4 min. For the subsequent FLA step, flash energies ranging from 34 up to 62 J/cm² were applied for 20.0 ms. Because of the short pulse times, direct temperature measurements during the flash are not possible [19], but a change in FLA energy of 7 J/cm² corresponds to a change in surface temperature of ~ 100°C.

### 2.4     *Sample Characterization*

All **GIXRD** patterns reported in this work were recorded using a Philips X'Pert MRD system equipped with a CuK$_\alpha$ x-ray ($\lambda = 0.154$ nm) source. A schematic sketch of a GIXRD measurement of a nanocrystalline sample with all relevant angles is depicted in Figure 1 a). The gracing incidence angle $\omega$ was chosen to be 0.3° for our samples for maximal intensity. The random orientation of the grains in nanocrystalline samples ensures that for all lattice planes (*hkl*) a group of NC is oriented to the incident beam such that the Bragg condition is fulfilled. The powder pattern of the samples can be recorded by a $2\theta$ scan. Such a powder pattern of a $Si_{0.63}C_{0.37}$ sample is depicted in Figure 1 b). The two main peaks (Si(111), cyan and SiC(111) magenta) are fitted to estimate the mean grain size of Si and SiC NC from the FWHM of the fits by the Scherrer equation. Grain size determined by GIXRD always refers to crystalline domain sizes determination without structural defects as twins or stacking faults. For randomly oriented grains, the recorded patterns are independent of the $\varphi$ angle around the normal of the sample surface. The measurement range was chosen to be $10° \leq 2\theta \leq 90°$. Additionally, the $\varphi$ angle was varied between $0°$ and $360°$. The GIXRD analysis is described in greater detail in our previous work [12, 20].

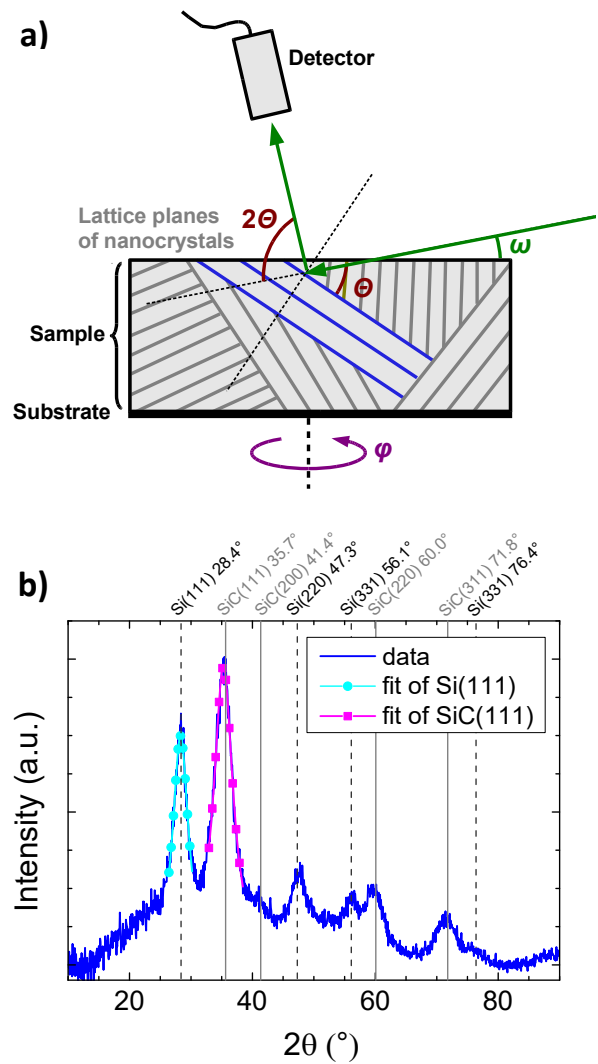

**Figure 1: a) Sketch of the beam path of a GIXRD measurement. b) A GIXRD pattern of an annealed Si$_{0.63}$C$_{0.37}$ sample.**

**FTIR** spectroscopy was conducted using a Bruker IFS 113v instrument over the range of 400 cm⁻¹ to 4000 cm⁻¹ with 6 cm⁻¹ resolution. The two most dominant IR modes in this range are the Si-C stretching mode $\nu_{Si-C}$ around 800 cm⁻¹ ([21] and references therein) and the Si-H stretching mode $\nu_{Si-H}$ between 2000 and 2200 cm⁻¹ [22, 23]. The latter allows for the observation of H effusion during annealing. The former can be used to assess Si-C crystallization by two different methods: The first method is based on the Si-C peak shift to higher wavenumbers with increasing Si-C crystallinity and provides qualitative insight into the crystallization process [7]. The second method involves quantifying the Si-C crystalline fraction by fitting the Si-C mode with a Lorentzian L($\nu$) and a Gaussian G1($\nu$) peak. This fitting is exemplified in Figure 2 for a $Si_{0.77}C_{0.23}$ layer flashed for 20 ms at 47 J/cm². Because L($\nu$) represents the crystalline Si-C bonds whereas G($\nu$) corresponds to the amorphous Si-C network, the relation of the peak areas yields an estimate for the Si-C crystalline fraction. This method is widely used in the literature [6, 24, 25] and was checked for applicability as well as described in detail for our measurements in a previous study [12]. To fit the Si-C mode well, one Lorentzian (L($\nu$)) and two Gaussian peaks (G1($\nu$) and G2($\nu$)) were used. The data treatment is exactly the same as that described in our last paper [12] with one exception: Therein we assigned G2($\nu$) the peak 850 cm$^{-1}$ to a second amorphous Si-C mode. Further measurements showed an increasing G2($\nu$) mode with increasing temperature, disproving the a-SiC assignment. However, we know from secondary ion mass spectroscopy (SIMS) measurements that there is up to $3\times10^{21}$ at/cm$^3$ nitrogen atoms in our as-dep layers [17]. Therefore, we assign G2($\nu$) to the asymmetric N-Si$_3$ stretching mode (830 – 890 cm$^{-1}$ [26, 27]) and verify this assignment by estimating the Si-N bond density:

$$N(\text{Si} - \text{N}) = K_{\text{Si}-\text{N}} \int_{\nu_1}^{\nu_2} \frac{\mu(\nu)}{\nu} d\nu \approx \frac{K_{\text{Si}-\text{N}}}{\nu_0} \int_{\nu_1}^{\nu_2} \mu(\nu) d\nu \text{ [28, 29].}$$

where $\int_{\nu_1}^{\nu_2} \mu(\nu) d\nu$ is the area and $\nu_0$ the central wavenumber of the G2($\nu$) mode. With $K_{\text{Si}-\text{N}} = 2\times10^{19}$ cm$^{-2}$ [27] and $\nu_0 = 850$ cm$^{-1}$, we obtain $N(\text{Si} - \text{N}) = 1.2\times10^{22}$ bonds/cm$^3$ for the G2($\nu$) mode in Figure 2, which corresponds to a nitrogen concentration of $4\times10^{21}$ at/cm$^3$, assuming threefold coordination. This number matches surprisingly well with the SIMS value and supports the assignment of G2($\nu$), which is thus not taken into account for the evaluation of Si-C crystallinity.

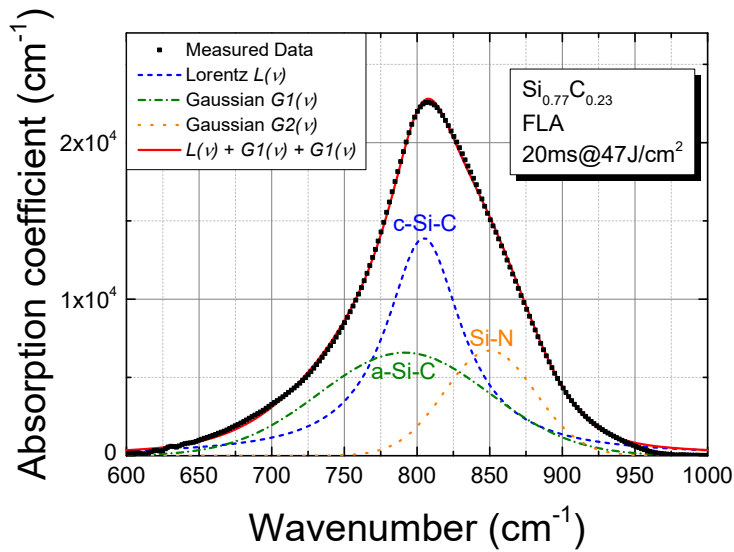

**Figure 2: Si-C stretching mode of an exemplary FTIR spectrum fitted with one Lorentzian L($\nu$) and two Gaussian peaks (G1($\nu$) and G2($\nu$)). L($\nu$) represents the crystalline and G1($\nu$) the amorphous part of the Si-C network, whereas G2($\nu$) is attributed to the asymmetric Si-N stretching vibration.**

**SEM** images were acquired using a Hitachi SU70 or a Zeiss AURIGA 60 FIB-SEM tool, both equipped with a Schottky emitter, operated at 5 kV accelerating voltage and a working distance of approximately 5 mm.

All **ellipsometry** data were acquired with a Woollam M-2000 spectroscopic ellipsometer and treated using the software WVASE®.

### III)   RESULTS

### 3.1   FLA parameter selection

### 3.1.1   Si content

We found that successful FLA for our samples was only possible for layers with a Si content of 77% and

63% but not 50%. The $Si_{0.50}C_{0.50}$ FLA samples, which had been treated with a preanneal step of

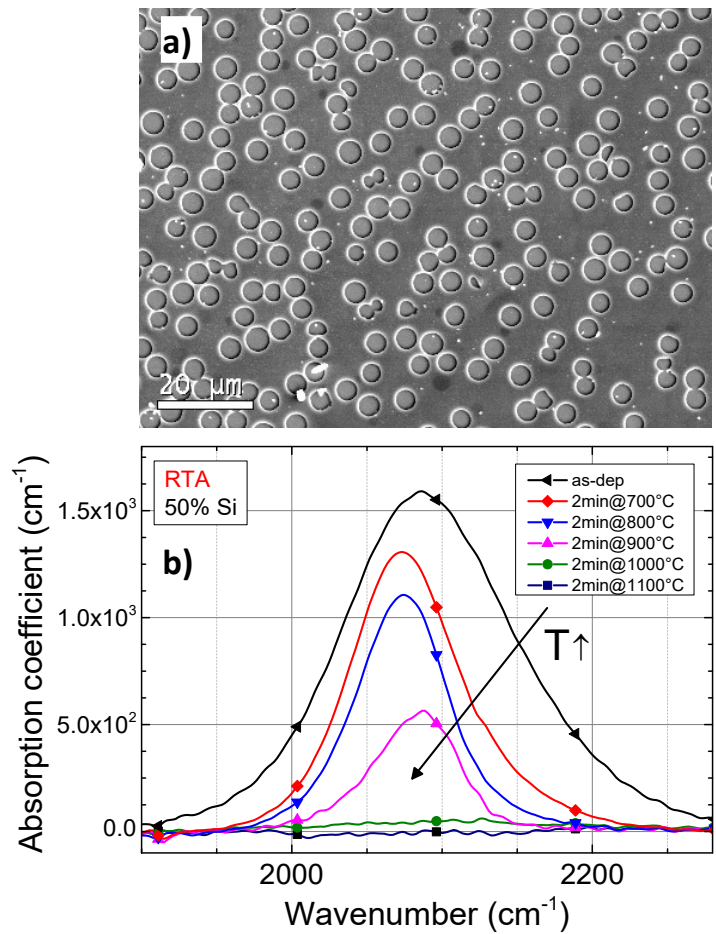

Figure 3: a) SEM image shows strong blistering (circular holes) of a $Si_{0.50}C_{0.50}$ film after FLA. b) Development of the FTIR Si-H mode with increasing annealing temperatures for the $Si_{0.50}C_{0.50}$ film. The latter shows that 700°C is not sufficiently hot for complete H effusion: After 2min@700°C, there is still a significant Si-H peak (red).

5min@700°C and flashed with energies between 28 and 47 J/cm², all showed a white haze on the surface.

SEM images (see Figure 3 a)) show that the origin of the white haze is strong blistering of the layers

originating from the explosive effusion of hydrogen (H) during the flash. Clearly, the preanneal step was

not hot and/or long enough for complete H effusion. This reasoning can be substantiated by the RTA

treatment of $Si_{0.50}C_{0.50}$ samples and by evaluating the corresponding FTIR measurements (Figure 3 b)). The Si-H vibration mode in the range of $2000\ldots2200$ cm$^{-1}$ reveals that after 2min@900°C (magenta) there was still a considerable number of Si-H bonds in the layers. Even after FLA, Si-H bonds were still visible in the FTIR spectra of $Si_{0.50}C_{0.50}$ samples (not shown here). Because the hottest and longest preanneal that is possible in the FLA setup is 2min@800°C, it was not possible to drive out all the hydrogen before FLA to avoid blistering of the stoichiometric $Si_{0.50}C_{0.50}$ samples. Therefore, no further investigation of $Si_{0.50}C_{0.50}$ samples was conducted in this work.

### 3.1.2 FLA energy

A wide range of flash energies were applied to $Si_{0.63}C_{0.37}$ and $Si_{0.77}C_{0.23}$ films. Because it was not clear what temperatures were achieved at each energy, it was necessary to check whether solid-phase crystallization was in fact taking place or whether some flash energies melted or otherwise damaged the samples. For flash energies above 40 and 47 J/cm² for $Si_{0.63}C_{0.37}$ and $Si_{0.77}C_{0.23}$, respectively, the GIXRD patterns retained their general shape, but the intensities of reflexes appeared to depend on the wafer orientation corresponding to the $\varphi$-angle, as shown in Figure 4 for the Si(111)@28.4° reflex.

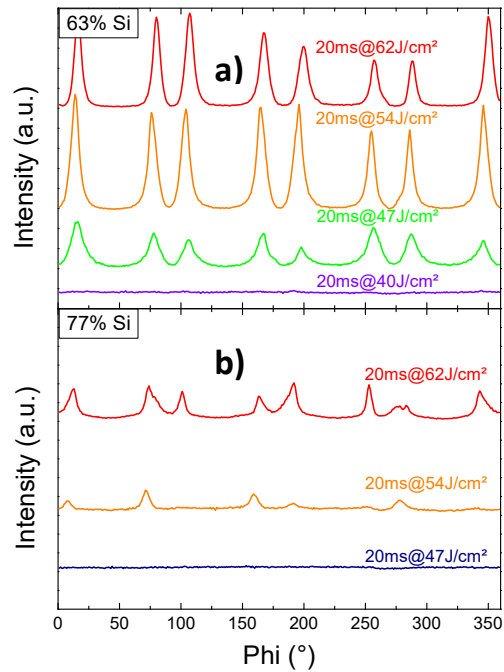

**Figure 4: The $\varphi$-orientation dependence of the Si(111) peak at 2θ = 28.4° with increasing FLA energy is shown for $Si_{0.63}C_{0.37}$ (a) and $Si_{0.77}C_{0.23}$ (b) layers. It is evident that for $Si_{0.63}C_{0.37}$ samples, the $\varphi$-dependency starts at lower flash energies and is stronger than for $Si_{0.77}C_{0.23}$ samples.**

$Si_{0.63}C_{0.37}$ films (Figure 4 a)) showed 4-fold symmetry, each with two peaks, which increased in intensity for 54 J/cm² and decreased slightly in intensity for 62 J/cm². In $Si_{0.77}C_{0.23}$ layers (Figure 4 b)), the $\varphi$-dependence only began at 54 J/cm² and increased for 62 J/cm². In contrast, the intensity of the SiC reflexes showed no dependence on sample orientation in either sample. Thus, at a certain flash energy the orientation of the Si NC was no longer randomly distributed, whereas the SiC grain orientation remained random for all energies. The evaluation of Si and SiC grain sizes from GIXRD spectra acquired at different sample

orientations shows that an increase in Si(111) peak intensity at a given orientation is correlated with an

increase in Si NC diameter in that particular direction. This behavior is not surprising because the GIXRD

intensity is in general positively correlated with the number of scattering centers [30] and hence with the size of crystalline clusters. This relationship is illustrated in Figure 5 a) for $Si_{0.63}C_{0.37}$ and in Figure 5 b) for

$Si_{0.77}C_{0.23}$.    The    connected    data    points    represent    the    GIXRD    measurements    in    the

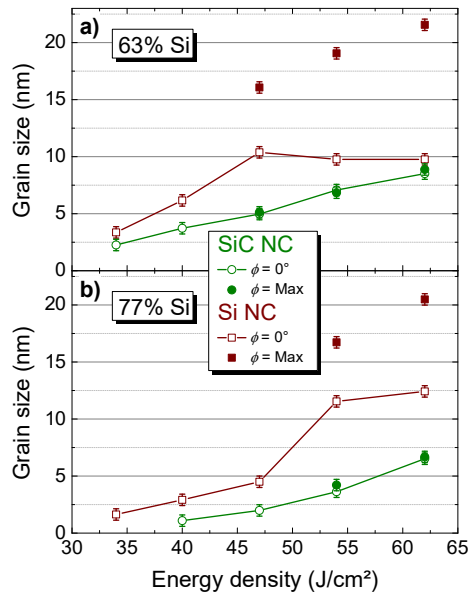

Figure 5: SiC (green circles) and Si (brown squares) grain sizes estimated from the Si(111) and SiC(111) reflex at $2\vartheta = 28.4°$ and at $2\vartheta = 31.7°$, respectively for $Si_{0.63}C_{0.37}$ (a) and $Si_{0.77}C_{0.23}$ (b). The straight lines connect the data points of measurements performed in the standard sample orientation referred to as $\varphi = 0°$. The additional grain sizes at high energies were calculated from diffraction patterns taken at $\varphi$-angles of maximal intensity of the Si(111) reflex at 28.4°. The Si NC size is a function of both, the flash energy and the sample orientation. The SiC NC size remains independent of the sample orientation even for high flash energies, but a large difference between the $Si_{0.63}C_{0.37}$ and the $Si_{0.77}C_{0.23}$ samples can be observed.

standard sample orientation, referred to as $\varphi = 0°$, whereas the additional grain sizes (filled symbols)

correspond to the spectra captured at the maximal intensity of the Si(111)@28.4° reflex. The largest

variation in Si NC grain size (brown squares) was observed for 63% Si samples flashed for 20 ms at 62 J/cm²

and ranged from $(9.8 \pm 0.5)$ nm to $(21.6 \pm 0.5)$ nm. The SiC NC grain size (green circles) did not depend on

sample orientation and increased continuously with the flash energy for layers with Si contents of 63% and

77%. The $\varphi$-dependence of the Si grain size and the very large NC suggest that at high energies, a different crystallization mechanism takes place, perhaps via the Si liquid phase ($T_m$(Si) = 1414°C [31]). As our aim is the solid-phase crystallization of Si NC, we forgo further investigation of the high-energy FLA samples.

In the following, we consider only samples with orientation-independent GIXRD spectra and establish a maximal suitable FLA energy of 47 J/cm² for $Si_{0.77}C_{0.23}$ and 40 J/cm² for $Si_{0.63}C_{0.37}$.

### 3.2 Formation of nanocrystals by RTA and FLA

#### 3.2.1 GIXRD

The nanostructure of samples annealed by RTA and FLA under conditions suitable for solid-phase crystallization was studied by GIXRD.

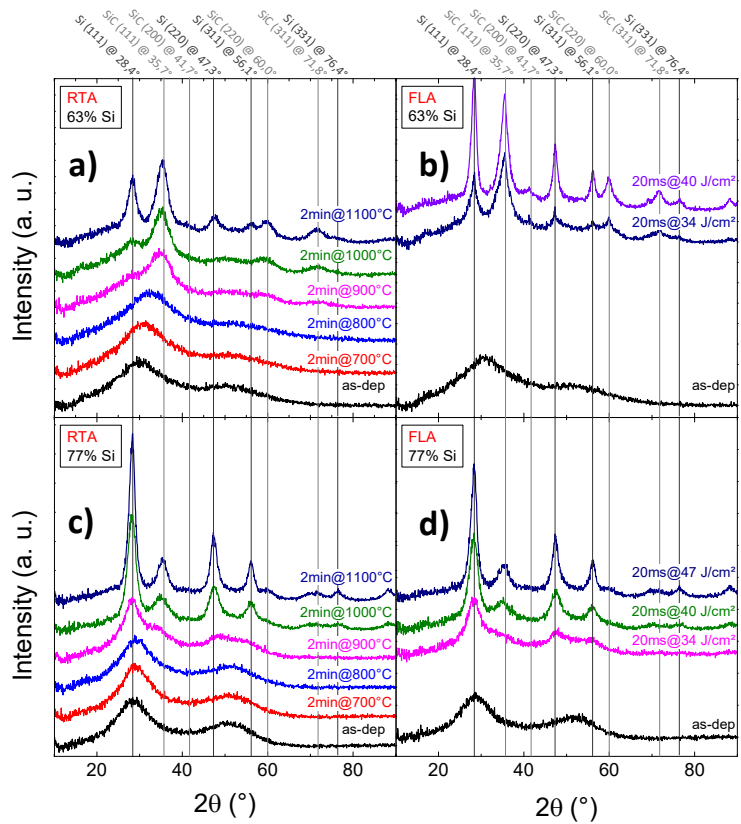

**Figure 6:** GIXRD patterns of RTA (a) and c)) and FLA (b) and d)) samples with a Si content of 63% (a) and b)) and 77% (c) and d)). To compare flash energies and annealing temperatures, the GIXRD patterns of FLA samples were plotted in line with GIXRD patterns of RTA samples with a similar shape.

As previously mentioned, the FLA process is characterized by the flash energy and not by the sample temperature. To compare the FLA results with the results obtained from RTA-processed samples, we gathered GIXRD results based on RTA temperature data and plotted the GIXRD results based on FLA energy data in line with the RTA data they most resemble. The final results are presented in Figure 6. All FLA samples were subjected to a preanneal step of either 2min@800°C (63% Si content, Figure 6 b)) or

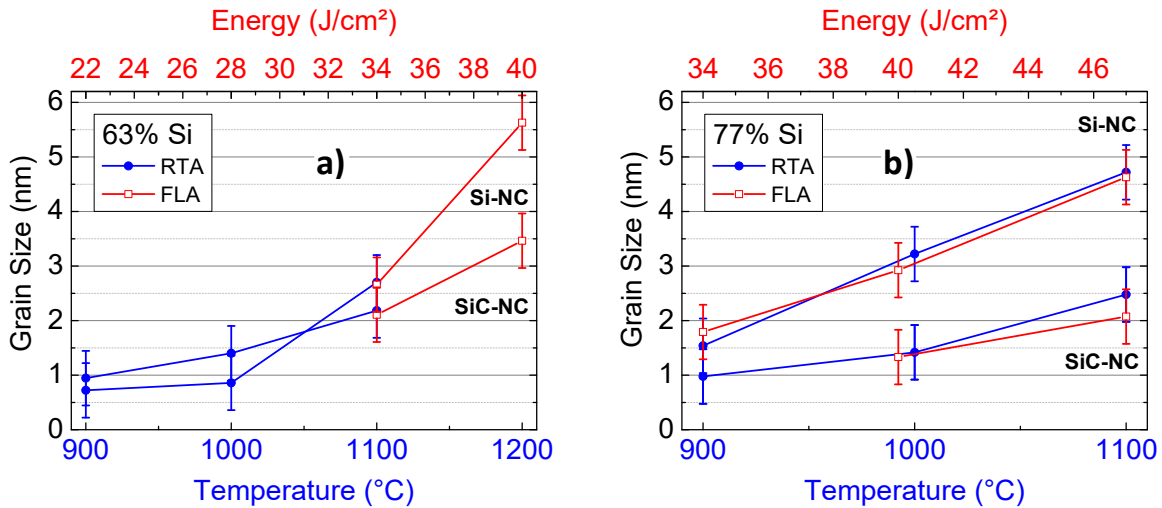

Figure 7: Grain sizes calculated from the GIXRD patterns in Figure 6 for $Si_{0.63}C_{0.37}$ (a) and $Si_{0.77}C_{0.23}$ (b). The correspondence between RTA temperatures and FLA energies is the same as that in Figure 6. Si NC size is observed to be a function of both Si content and annealing temperature, whereas SiC NC size is independent of film composition and depends on annealing temperature and the corresponding flash energy.

4min@700°C (77% Si content, Figure 6 d)) before FLA. At a Si content of 63%, the highest RTA temperature of 1100°C yielded data similar to that obtained at the lowest FLA energy of 34 J/cm². The situation was different for a Si content of 77%, at which a flash energy of 34 J/cm² produced a GIXRD pattern similar to that observed for RTA at 900°C. The data obtained from $Si_{0.77}C_{0.23}$ samples exposed to FLA at 40 and 47 J/cm² can be compared with those obtained from films that underwent RTA at 1000 and 1100°C, respectively. The average grain sizes of Si and SiC NC were calculated from the GIXRD spectra and are plotted as a function of the RTA temperature and the FLA energy in Figure 7. Therein, the attribution of the flash energies to a certain RTA temperature corresponds to the attribution of the spectra in Figure 6. The results confirm our assumption regarding the GIXRD spectra: In the $Si_{0.63}C_{0.37}$ samples (Figure 7 a)), the hottest RTA temperature of 1100°C led to the same NC grain sizes (approximately 2.2 and 2.6 nm for SiC and Si NC, respectively) produced by FLA at the lowest energy of 34 J/cm². For higher flash energies, the grains continued to grow to $3.5 \pm 0.5$ nm in diameter for SiC NC and $5.6 \pm 0.5$ nm for Si NC, larger than all Si NC we ever produced in these films with RTA or a furnace anneal in prior studies [12, 20]. The grain sizes in the $Si_{0.77}C_{0.23}$ samples (Figure 7 b)) confirm that the flash with 34 J/cm² led to NC comparable to those formed by RTA at 900°C and the increase to 47 J/cm² corresponds to an increase in temperature to 1100°C. The results also confirm findings reported in our previous work [12], which showed that the Si NC size is a function of the Si content in samples whereas the SiC NC size depends only on the annealing temperature. The FLA energy that evokes the same Si NC sizes as RTA at a certain temperature increases with Si content. For example, Figure 6 and Figure 7 show that 34 J/cm² produced larger NC in $Si_{0.63}C_{0.37}$ than in $Si_{0.77}C_{0.23}$, which means that the crystallization process during FLA is favored in films containing less Si. This reasoning is counterintuitive because Si crystallization would be expected to be easier if more Si is present. The same trend was observed for FLA at higher energies, as will be discussed in greater detail in the discussion section.

### 3.2.2 FTIR

Figure 8 shows the FTIR spectra of all $Si_{0.77}C_{0.23}$ and $Si_{0.63}C_{0.37}$ samples on Si substrates treated by RTA (Figure 8 a) and c)) and low-energy FLA (Figure 8 b) and d)). The dominant Si-C stretching vibration around 800 cm$^{-1}$ is depicted in the main panels, whereas the insets show the Si-H stretching vibration at approximately 2000 cm$^{-1}$. It is important to note that neither for RTA nor for FLA could the formation of Si-O bonds be observed because no peak around 1000 cm$^{-1}$ appeared at higher temperatures.

The area of the Si-C mode increased and the peaks shifted to higher wavenumbers as the annealing temperatures as well as the flash energies increased. The latter is associated with the SiC crystallization process, whereas the former corresponds to a growing Si-C bond density in the samples. As expected based on the film composition, the $Si_{0.77}C_{0.23}$ samples showed smaller peak areas and hence lower Si-C bond densities than the $Si_{0.63}C_{0.37}$ samples under comparable annealing conditions. Another difference caused by the Si content was the temperature of complete H effusion $T_{out}$. Comparing the insets of Figure 8 a) and c) reveals that $T_{out} \geq 800°C$ in the $Si_{0.77}C_{0.23}$ samples, whereas for the $Si_{0.63}C_{0.37}$ samples $T_{out} \geq 900°C$. As mentioned previously (Figure 3 b)), $T_{out} \sim 900°C$ in $Si_{0.50}C_{0.50}$ RTA samples. This trend appears to be the same for FLA; indeed, the insets of Figure 8 b) and d) show that the area of the Si-H mode was larger for $Si_{0.63}C_{0.37}$ preannealed at 800°C than for $Si_{0.77}C_{0.23}$ preannealed at 700°C. As discussed in section 3.1, no suitable $Si_{0.50}C_{0.50}$ FLA samples exist. The dependence of $T_{out}$ on Si content can be explained by the differences in back-bonding. With increasing Si content, the nearest neighbors of Si atoms to which H is bonded are increasingly other Si atoms, not C atoms. Because Si atoms show a lower electronegativity than C atoms, the H atoms are less strongly bonded with increasing Si content and can effuse at lower temperatures [32, 33]. This explanation is supported by the fact that the Si-H$_n$ mode in as-dep samples shifts to lower wavenumbers with increasing Si content (see Figure 3 b) and inset of Figure 8 a) and c)), which corresponds to a weakening of the corresponding bond.

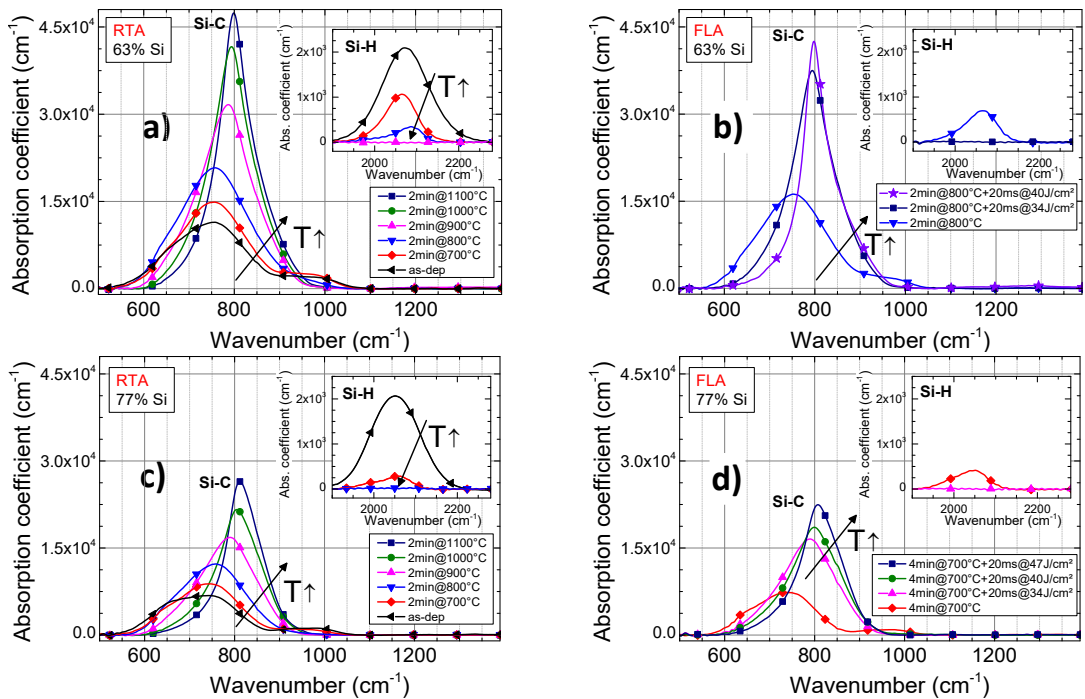

**Figure 8:** FTIR spectra of RTA (a) and (c) and FLA (b) and (d) samples with a Si contents of 63% (a) and (b) and 77% (c) and (d). Whereas the Si-C mode increases with temperature und flash energy, the Si-H vibration depicted in the insets decreases.

The contribution of H from C-H bonds to H effusion could not be examined in this work because the C-H FTIR bonds were too weak to analyze.

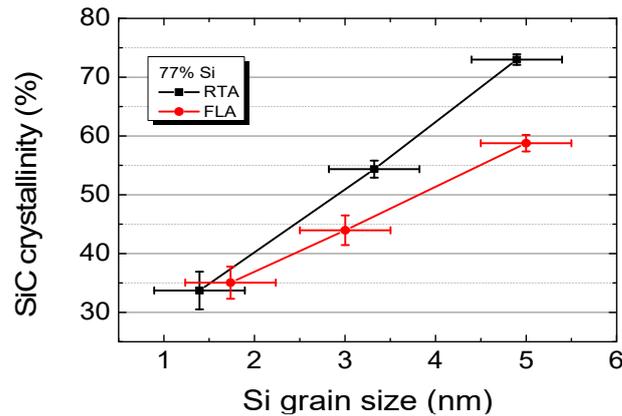

**Figure 9: The SiC crystallinity determined from FTIR spectra as a function of the Si NC size derived from GIXRD patterns show that FLA leads to less SiC crystallinity than RTA for a given Si NC size.**

Whereas the overall trends in FTIR spectra of FLA and RTA samples are similar, a careful analysis (described in Section 2.4) of the SiC crystallization in $Si_{0.77}C_{0.23}$ samples reveals a slight difference, as depicted in Figure 9. For Si grain sizes larger than 2 nm (as determined by GIXRD), we observed significantly lower SiC crystallinity in FLA samples compared with that observed in RTA samples for a given Si grain size. The difference in SiC crystallinity increased with increasing Si NC size. This trend will be discussed in the next section.

IV)    DISCUSSION

### 4.1 Reduced SiC crystallinity in FLA samples

Figure 9 shows a decrease in SiC crystallinity in FLA samples relative to that of RTA samples for the same Si NC size. This result is remarkable because a reduction in SiC crystallinity promises a defect reduction in samples [8]. Because Figure 7 b) shows that SiC NC size remained the same within experimental error for a given Si NC size, irrespective of whether FLA or RTA was used, the reduced SiC crystallinity for FLA must have resulted from a lower number of SiC NC in FLA compared with that in RTA. This finding can be explained with the help of classical crystallization theory in terms of nucleation and crystal growth. Theory states that the temperature $T_N$ at which maximal nucleation takes place is lower than the temperature of maximum crystal growth $T_G$ [34]. Near the phase transition temperature $T_T$, which is the melting point for Si ($T_T^{Si} = 1414°C$ [31]) and the decomposition point for SiC ($T_T^{SiC} > 2300°C$), there is rapid crystal growth but a marginal nucleation rate. The RTA samples shown in Figure 9 were annealed between 900°C and 1100°C, and the FLA samples were exposed to comparable annealing conditions, as demonstrated by the consistency of the structural measurements shown in Figure 6 and Figure 8. Thus, in RTA and FLA, $T \ll T_T^{SiC}$ such that both nucleation and growth of SiC NC occur throughout the RTA and FLA processes. The larger number of SiC NC in RTA samples may therefore be due to a longer plateau time or a longer heating ramp. The plateau time was 20 ms for FLA and 2 min for RTA, and the ramp-up time was 22 s for RTA at 1100°C. In the case of FLA, the heating of the samples was limited only by the thermal properties of the layer. The thermal diffusion length $L_D$ is the penetration depth of heat into a material with thermal diffusivity $\alpha$ during a pulse duration $t$:

$$L_D = \sqrt{\alpha t}.$$

The thermal diffusivity is given by

$$\alpha = \frac{k}{\delta \, c_p}$$

The other parameters are the thermal conductivity $k$, the density of the material $\delta$ and the specific heat capacity $c_p$:

**Table 1: Literature values for the determination of the thermal diffusivity $\alpha$ and the thermal diffusion length $L_D$ in Si and SiC.**

| Material | $T$ | $k$ (W/ cm K) | $c_p$ (J/ g K) | $\delta$ (g/ cm³) | Ref. | $\alpha = k/(\delta \cdot c_p)$ (cm²/ s) | $L_D = \sqrt{\alpha t}$ (µm) for $t$ = 20 ms | $t = L_D{}^2/\alpha$ (ms) for $L_D$ = 200 nm |
|---|---|---|---|---|---|---|---|---|
| **a-Si** | 27°C | 0.007 | 0.77 | 2.30 | [35] | 0.0040 | 89 | 0.1012 |
| | 1000°C | 0.011 | 1.07 | 2.30 | [35] | 0.0045 | 95 | 0.0895 |
| | 1145°C | 0.013 | 1.12 | 2.30 | [35] | 0.0050 | 100 | 0.0793 |
| **poly-Si** | 27°C | 0.16 | as c-Si | as c-Si | [36] | 0.1 [36] | | |
| **c-Si** | 27°C | 1.56 | 0.71 | 2.33 | [35] | 0.9430 | 1373 | 0.0004 |
| | 1000°C | 0.25 | 0.96 | 2.30 | [35] | 0.1132 | 476 | 0.0035 |
| | 1414°C | 0.22 | 1.03 | 2.29 | [35] | 0.0933 | 432 | 0.0043 |
| **a-SiC:H** | 27°C | | | 1.51 - 2.15 | [37] | | | |
| **a-SiC** | 27°C | 1.3 - 1.6 | | | [38] | | | |
| **poly-SiC** (with additives) | 27°C | 0.48 | 0.7 | | [39] | | | |
| | 1000°C | 0.27 | 1.2* | | [39] | | | |
| | 1500°C | 0.18* | | | [39] | | | |
| **c-SiC** | 27°C | 4.9 | 0.85 | 3.217* | [40, 41] | | | |
| | 130°C | | 1.04 | | [42] | | | |
| | 330°C | | 1.20 | | [42] | | | |
| | 730°C | | | | [42] | | | |

For our $Si_xC_{1-x}$ layers, these values are not known, but we can estimate $\alpha$ based on values reported for a-Si and SiC in the literature, as summarized in Table 1. For a-SiC:H, there is no complete dataset available, but a lower bound on its thermal diffusivity $\alpha_{SiC}^{min}$ can be estimated based on the values marked with a * in Table 1. The resulting $\alpha_{SiC}^{min}$ of 0.0047 cm²/s is higher than $\alpha_{a-Si}^{1000°C} = 0.0045$ cm²/s. Therefore, we can use $\alpha_{a-Si}^{1000°C}$ for a lower limit of the thermal diffusivity in the $Si_xC_{1-x}$ layers, which indicates that the heating time in FLA is less than 0.1 ms and therefore five orders of magnitude shorter than in the case of RTA. Consequently, compared with RTA, the FLA process leads to the formation of fewer SiC grains during ramping-up *and* during the temperature plateau.

### *4.2 Influence of Si-content on crystallization behavior:*

As shown in Figure 6 and Figure 5, a certain FLA energy leads to the formation of larger Si and SiC NC in $Si_{0.63}C_{0.37}$ than in $Si_{0.77}C_{0.23}$ layers. Possible explanations include (i) better absorption in the $Si_{0.63}C_{0.37}$ than in the $Si_{0.77}C_{0.23}$ layers (although this would be counterintuitive); (ii) a strong effect of the preanneal step, which was 2min@800°C for $Si_{0.63}C_{0.37}$ and 4min@700°C for $Si_{0.77}C_{0.23}$ samples; and (iii) the effect of the amount of hydrogen remaining after the preanneal step, which depends on the Si content. Each possibility is examined in turn.

(i): We verified the absorption of the samples by modeling $A_{L-i}(\lambda) = 1-R_i(\lambda)-T_i(\lambda)$ (i = 63, 77) based on the dielectric function $\varepsilon_i(E)$ for preannealed $Si_{0.63}C_{0.37}$ and $Si_{0.77}C_{0.23}$ layers on Si substrate. $\varepsilon_i(E)$ were obtained

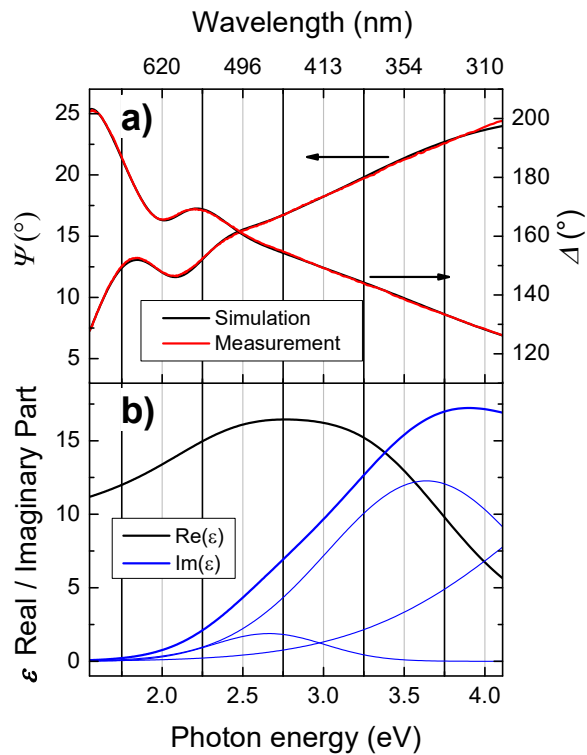

**Figure 10: a) Ellipsometry data (red) and simulation (black) of a-$Si_{0.77}C_{0.23}$:H sample. b) Simulated imaginary part of the dielectric function $\varepsilon(E)$ (blue, thick) with three Gaussian contributions (blue, thin) and real part (black) derived by Kramers-Kronig relation.**

via spectroscopic ellipsometry. The simulation based data analysis was performed within the software WVASE® employing a parametric model of 3 oscillators, each contributing a Gaussian profile to the imaginary part of the dielectric function:

$$\varepsilon_{\mathrm{G}}(E) = A \exp\left(\frac{(E-E_0)^2}{4\ln 2\sigma^2}\right).$$

with $A$ being the dimensionless amplitude, $E_0$ the position of the Gauss peak and $\sigma$ the full width at half maximum. The contribution to the dielectric function real part is internally calculated via Kramers-Kronig relations. In Figure 10, the simulated and measured $\Psi$ and $\Delta$ spectra (Figure 10 a)), as well as the dielectric function results (Figure 10 b)) with the individual Gaussian contributions (lighter blue lines) are shown for the a-Si$_{0.77}$C$_{0.23}$:H sample. Results obtained with a Tauc-Lorentz model [43] were found to produce the same characteristic run of the dielectric function. However, in the low energy regime this model provided insufficient flexibility. These dielectric functions were then used to simulate $R_i(\lambda)$ and $T_i(\lambda)$ and thereby $A_{\mathrm{L}\text{-}i}(\lambda)$ with the help of the software CODE. Figure 11 a) shows the calculated absorption $A_{\mathrm{L}\text{-}i}(\lambda)$ for Si$_{0.63}$C$_{0.37}$ (blue) and Si$_{0.77}$C$_{0.23}$ (green) (left-hand ordinate) with the normalized intensity of the FLA Xe lamps $I(\lambda)$

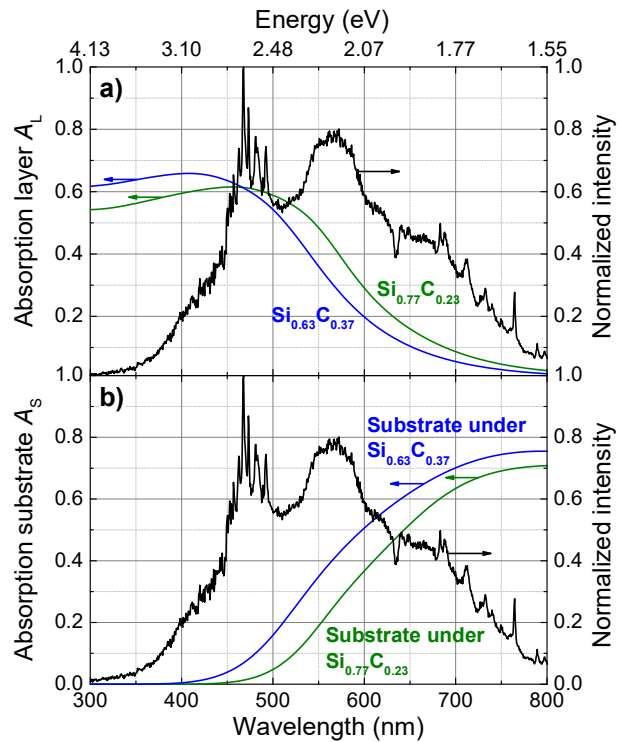

**Figure 11: The normalized emission spectrum of the Xe lamps (black) used for FLA is plotted with the absorption of the layers Si$_{0.77}$C$_{0.23}$ ($A_{\mathrm{L}\text{-}77}$, green) and Si$_{0.63}$C$_{0.37}$ ($A_{\mathrm{L}\text{-}63}$, blue) in a) and the absorption of the substrate under the Si$_{0.77}$C$_{0.23}$ ($A_{\mathrm{S}\text{-}77}$, green) and Si$_{0.63}$C$_{0.37}$ ($A_{\mathrm{S}\text{-}63}$, blue) layer in b).**

[18] (right-hand ordinate). Figure 11 b) presents the calculated $T_i(\lambda)$ for Si$_{0.63}$C$_{0.37}$ (blue) and Si$_{0.77}$C$_{0.23}$ (green) (left-hand ordinate), which correspond to the absorption in the Si substrate $A_{\mathrm{S}\text{-}i}(\lambda)$ under the Si$_{0.63}$C$_{0.37}$

and the $Si_{0.77}C_{0.23}$ layer ($T_i(\lambda) = A_{S-i}(\lambda)$). Figure 11 a) shows that $A_{L-77}(\lambda) > A_{L-63}(\lambda)$ in the spectral range of high intensity of the Xe lamps. Correspondingly, more light is transmitted and therefore absorbed in the Si substrate in the case of the $Si_{0.63}C_{0.37}$ sample ($A_{S-63}(\lambda) > A_{S-77}(\lambda)$). To quantify this observation, the fraction of $I(\lambda)$ absorbed in the layer was calculated by

$$\chi_{L-i} = \frac{\int A_{L-i}(\lambda) I(\lambda)\, d\lambda}{\int I(\lambda) d\lambda}$$

and in the substrate by

$$\chi_{S-i} = \frac{\int A_{S-i}(\lambda) I(\lambda)\, d\lambda}{\int I(\lambda) d\lambda}$$

to be $\chi_{L-63} = 32\%$ and $\chi_{L-77} = 37\%$ in the layers and $\chi_{S-63} = 39\%$ and $\chi_{S-77} = 30\%$ in the substrate. It follows that the absorption behavior of the layers alone cannot be the reason for the stronger crystallization in $Si_{0.63}C_{0.37}$ compared with that in $Si_{0.77}C_{0.23}$. However, the overall absorption $\chi_i = \chi_{L-i} + \chi_{S-i}$ is higher in the $Si_{0.63}C_{0.37}$ sample ($\chi_{63} = 71\%$) than in the a-$Si_{0.77}C_{0.23}$ sample ($\chi_{77} = 67\%$). Therefore, it is possible that the absorption in the substrate can act as a significant heat source for the layer on top. To verify this reasoning, we performed a further experiment with the same a-$Si_{0.63}C_{0.37}$:H and a-$Si_{0.77}C_{0.23}$:H layers deposited on quartz instead of Si substrate because quartz shows negligible absorption in the optical wavelength range. After a preanneal of 2min@800°C for a-$Si_{0.63}C_{0.37}$:H and 4min@700°C for a-$Si_{0.77}C_{0.23}$:H, an FLA of 20ms@47 J/cm² was performed for both samples. The Si NC size derived from GIXRD measurements was ($14.6 \pm 0.5$) nm for $Si_{0.63}C_{0.37}$ and ($18.6 \pm 0.5$) nm and thus significantly higher for $Si_{0.77}C_{0.23}$. This result agrees with our expectation of an increase in Si NC size with increasing Si content and suggests that strong absorption in the substrate promotes film crystallization.

(ii) Because $Si_{0.63}C_{0.37}$ was preannealed at higher temperatures than $Si_{0.77}C_{0.23}$, the formation of more crystallization seeds during preannealing in $Si_{0.63}C_{0.37}$ samples compared with $Si_{0.77}C_{0.23}$ could be the result. However, even if this was the case, this effect would only lead to an increased number of Si and SiC crystallites in $Si_{0.63}C_{0.37}$ samples compared with the number observed in $Si_{0.77}C_{0.23}$ samples, as discussed in the previous section, but would not explain the observed difference in NC sizes. Thus, the preanneal temperature is not the reason for the crystallization behavior.

(iii) As shown in the inset of Figure 8 b) and d), the Si-H vibrational mode of $Si_{0.63}C_{0.37}$ (blue) after preannealing is stronger than that in $Si_{0.77}C_{0.23}$ (red), indicating more remaining hydrogen in $Si_{0.63}C_{0.37}$ samples after preannealing than in $Si_{0.77}C_{0.23}$. This higher amount of hydrogen could result in enhanced Si diffusion in $Si_{0.63}C_{0.37}$ compared with that in $Si_{0.77}C_{0.23}$, leading to the formation of larger clusters at the same flash energy and therefore to larger Si NC in $Si_{0.63}C_{0.37}$ samples. Lanckmans *et al.* [44] observed a similar effect of H on the mobility of $Cu^+$ in PECVD deposited a-SiC:H layers.

We consider both increased hydrogen content after preannealing and substrate absorption of FLA light to be probable explanations for the unexpected improvement in Si crystallization with decreasing Si content.

V)        CONCLUSIONS

A comparison of $Si_xC_{1-x}$ layers with varying compositions ($x = 0.50$, $x = 0.63$ and $x = 0.77$) treated by RTA and by FLA reveals that the reduction from an annealing time of 2 min to 20 ms leads to surprisingly similar crystallization results, as evidenced by GIXRD and FTIR measurements. We showed that a preanneal step for H effusion is needed to avoid blistering during FLA, whereas for RTA the ramping-up time is sufficiently long for H effusion without blistering. The H effusion temperature increases with decreasing Si content in the layer and exceeds the maximum possible preanneal temperature in the FLA system for $Si_{0.50}C_{0.50}$ samples. Therefore, successful FLA was only possible for $Si_{0.63}C_{0.37}$ and $Si_{0.77}C_{0.23}$ samples.

For both $Si_{0.63}C_{0.37}$ and $Si_{0.77}C_{0.23}$ layers, the Si NC lose their randomly distributed orientation and grain size above a certain FLA energy (40 J/cm² and 47 J/cm², respectively), whereas the orientation of the SiC NC remain random for all flash energies. Because this Si grain orientation require highly mobile Si atoms, we suggest that a Si liquid phase is formed at high flash energies, and we take the onset of the Si grain orientation process as an upper energy limit for solid-phase crystallization by FLA.

The $Si_{0.63}C_{0.37}$ show larger Si and SiC NC sizes than the $Si_{0.77}C_{0.23}$ samples for the same FLA energy over the entire energy range. One possible explanation for this trend is the amount of hydrogen remaining in the layers after the preanneal step. This amount is higher for $Si_{0.63}C_{0.37}$ than for $Si_{0.77}C_{0.23}$ layers and could facilitate diffusion and therefore crystallization in the $Si_{0.63}C_{0.37}$ layers. Another explanation is the different absorption behavior of the $Si_{0.63}C_{0.37}$ and $Si_{0.77}C_{0.23}$ layers on Si substrates. Taking into account the absorption in both the layer and the substrate, our calculation results in higher absorption in the $Si_{0.63}C_{0.37}$:H sample ($\chi_{63} = °71\%$) than in the a-$Si_{0.77}C_{0.23}$:H sample ($\chi_{77} = °67\%$). This discrepancy could cause a higher sample temperature in $Si_{0.63}C_{0.37}$:H than in a-$Si_{0.77}C_{0.23}$:H samples and therefore the formation of larger NC.

A very promising observation is the reduction of SiC crystallization as a result of switching from RTA to FLA. This reduction was explained in terms of nucleation and crystal growth and led to the formation of fewer SiC NC during FLA compared with the number observed in RTA samples. FLA is a promising method for synthesizing Si NC embedded in less crystalline SiC, providing a route towards developing Si

NC in a-SiC, which are expected to be of higher electronic quality and thus enable the fabrication of better

Si NC solar cells.

VI)     ACKNOWLEDGEMENTS

The authors wish to thank Philipp Barth, Mira Kwiatkowska, Antonio Leimenstoll and Felix Schätzle for help with sample processing and Jutta Zielonka for SEM measurements. The research leading to these results has received funding from the Protestant Academic Foundation Villigst, Germany.